\documentclass{iopjournal}
\usepackage{amsmath}
\usepackage{multirow}
\usepackage{amssymb}
\usepackage{graphicx}
\usepackage{booktabs}
\usepackage[T1]{fontenc}
\usepackage{lmodern}
\usepackage{comment}
\usepackage{tikz}
\usetikzlibrary{arrows.meta}
\usetikzlibrary{arrows.meta, calc}

\begin{document}
\hbadness=10000

\articletype{Paper}

\title{Global Structure in Learned Latent Representations of Confusion-Limited LISA Data}

\author{Jericho Cain$^{1,*}$\orcid{0000-0003-4731-9142}}

\affil{$^1$Physics Department, Portland State University, Portland, OR, USA}

\affil{$^*$Author to whom any correspondence should be addressed.}

\email{jcain@pdx.edu}


\begin{abstract}
Machine learning methods in gravitational wave data analyses depend on the choice of representation and on how structure within that representation is used. Building on previous work using continuous wavelet transform (CWT) autoencoder representations for confusion-limited LISA simulation, we investigate whether source resolvability information is better characterized by local latent geometry or by global latent density. We study this question in a controlled benchmark with data generation and preprocessing held fixed. Using CWT representations of synthetic confusion-limited LISA segments, we compare geometry based one-class scoring with likelihood-based latent models along with their morphology augmented variants. Likelihood-based scoring consistently outperforms local manifold-distance methods across three independent seeds, achieving ROC-AUC \(0.8555 \pm 0.0181\) and PR-AUC \(0.9219\pm 0.0118\), compared with ROC-AUC \(0.7663\pm 0.0450\) and PR-AUC \(0.8667\pm 0.0255\) for the geometry baseline. These results suggest that resolvability information in learned latent representations is not fully captured by local latent geometry but instead reflects global properties of the latent distribution. More broadly, this work contributes to representation-aware methods for confusion-foreground characterization in LISA and motivates future studies of coordinate invariance and intrinsic geometry in learned latent spaces.
\end{abstract}

\section{Introduction}
\begin{figure}[htbp]
    \centering
    \includegraphics[width=0.95\linewidth]{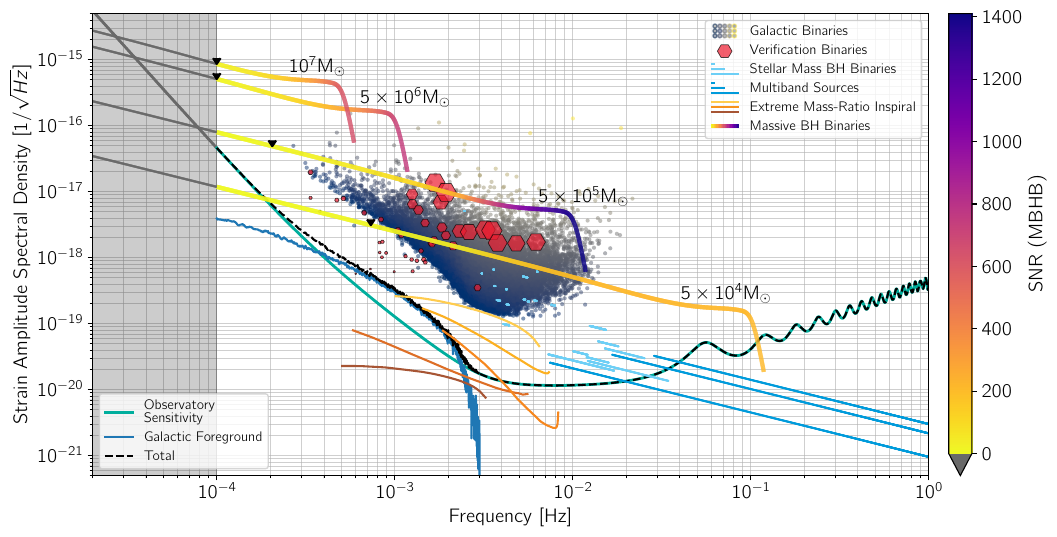}
    \caption{
Primary LISA source classes in the frequency--strain plane.
The instrumental sensitivity, unresolved galactic confusion
foreground, and total sensitivity are shown together with
representative source populations. Adapted from
\cite{colpi2024lisadefinitionstudyreport}.
}
\label{fig:lisa_source_classes}
\end{figure}

The mHz gravitational-wave band targeted by LISA is expected to be crowded with galactic compact binaries. At low frequencies, many of these systems are unresolved and form a confusion-limited foreground; at higher frequencies or amplitudes, a smaller subset becomes individually resolvable \cite{Nelemans2001,Timpano2006}. This foreground is scientifically valuable but it is also a practical obstacle in that unresolved structure can hide or distort weaker sources \cite{Crowder2007}. Figure~\ref{fig:lisa_source_classes} illustrates the principal LISA source populations together with the instrumental sensitivity and unresolved galactic foreground expected for the mission.

Foreground modeling and source-separation strategy are therefore not optional implementation details. They define the detection setting itself. Earlier studies showed that identifying and subtracting bright systems can materially change residual foreground statistics and therefore the effective detection landscape \cite{Timpano2006,Crowder2007}. Sensitivity-curve assumptions and mission-level requirements add another layer of realism to this problem \cite{Robson2019,LISASciRD}.

In this paper we focus on one-class scoring in latent feature space, motivated by the regime where anomalies are rare, diverse, or poorly labeled. Instead of training a closed-set classifier, we model the dominant foreground class and rank candidates by how atypical they are. Using several scoring rules, with data generation and preprocessing held fixed, we investigate which scoring family better separates resolvable sources from the confusion foreground.

We compare two approaches: local geometric deviation in latent space (manifold-style scoring) and explicit latent density modeling (likelihood-style scoring). Geometry-based scores are intuitive for autoencoder latent vectors, but dense foreground structure can make local distance less reliable than expected. Likelihood methods can better capture global foreground structure, though they introduce their own modeling choices. By evaluating these approaches under identical conditions and across multiple seeds, we show that likelihood-based latent scoring is the more robust choice in this setting, and we provide reproducible artifacts to support follow-up methodological work.

We train a CWT autoencoder solely on confusion-limited data until representation has been learned. Using that latent representation we investigate whether source resolvability information appear primarily through local latent geometry or through global latent distribution. The goal here is not to replace matched filtering, but to construct low-dimensional latent vectors that preserve the intrinsic geometry of the waveform manifold. 

If the data lives on a lower dimensional manifold, then computations using that data are carrying around a lot of redundancy. Learning meaningful coordinates could let us do the same physics with fewer degrees of freedom and therefore less computation. Such coordinates may support visualization, clustering, anomaly detection, source separation, accelerated inference, and other downstream analyses. 
\section{Related Work}

Classical gravitational-wave searches were built around two strong ideas: matched filtering for modeled sources, and coherent burst methods for less-modeled transients. Template placement theory remains the backbone of compact-binary searches \cite{Owen1996}. In parallel, multiresolution and coherent burst pipelines established a robust time--frequency route for non-template analyses \cite{Chatterji2004,Klimenko2008}.

The LISA setting raises a different challenge. In the milli-Hz band, unresolved galactic binaries are expected to produce a confusion foreground over a nontrivial frequency range \cite{Nelemans2001,Timpano2006}. The confusion foreground is not a side issue; it is part of the operating environment for source characterization \cite{Crowder2007}. Mission and sensitivity studies reinforce this framing, both in instrument-level requirements and in practical sensitivity construction \cite{AmaroSeoane2017,LISASciRD}. Updated sensitivity-curve treatments make the same point in analysis form \cite{Cornish2017,Robson2019}.

Machine learning has already shown value in gravitational-wave pipelines. Deep models have been used for fast detection and parameter-estimation-adjacent tasks in ground-based data \cite{Gabbard2018,George2018}. Other work has focused on detector-noise structure and data-quality classification \cite{Powell2015}. There are also modern examples of unsupervised or source-agnostic strategies, including recurrent autoencoder approaches and newer anomaly-detection studies \cite{Moreno2022,Fayad2024}. These results are promising, but most papers evaluate one family at a time.

Our recent papers are the direct lead-in. \cite{Cain2026} showed that CWT-based autoencoder representations can support template-free scoring in realistic LIGO conditions. Our recent work \cite{Cain2025} introduced CWT-based autoencoder representations for confusion-limited source separation and evaluated local latent geometry-based one-class scoring in that setting. In that work, we found manifold-learning, in this paper called the geometric baseline, significantly outperformed the reconstruction error of a standard autoencoder through augmentation with off-manifold distance. This showed that the latent space had useful structure that could be used to represent waveforms in continuous wavelet space.

This paper addresses which latent space scoring rule makes the most use of that structure.  That question has a clear precedent in the anomaly-detection literature. One branch focuses on support estimation and one-class decision boundaries \cite{Schoelkopf2001}. Another branch models density explicitly, typically with mixture models fitted by expectation-maximization (EM) or with nonparametric estimators \cite{Dempster1977,Parzen1962}. In gravitational-wave analyses, these branches are usually evaluated in different data setups, with different preprocessing and operating assumptions, making direct comparison difficult. Here the latent scoring rules are presented with an identical latent space structure generated through training on identical confusion foreground data, and presented with identical test data. The outcome of these results allow us to draw conclusions about whether the resolvability information in learned latent representations is fully captured by local latent geometry or global structure.

\section{Data and Experimental Protocol}

\subsection*{Synthetic benchmark design}

This study uses a controlled synthetic benchmark to compare one-class scoring strategies in a confusion-limited LISA setting. The design follows the confusion-foreground framing used in prior LISA studies \cite{Nelemans2001,Timpano2006,Crowder2007} and is aligned with mission-level sensitivity context \cite{Robson2019,LISASciRD}. In this section we summarize the same data generation used in our prior study \cite{Cain2025}.

LISA will observe for years but analysis methods need not consume year long arrays as indivisible objects. Long streams are commonly analyzed through finite windows or overlapping segments. We generate windows in the time domain with duration \(T=3600\) s and sampling rate \(f_s=1\) Hz, yielding 3600 samples per segment. This is not a claim that all LISA inference should be performed on one-hour data products. This choice of duration provides a computationally tractable benchmark with fixed input dimensionality while retaining time-frequency morphology in the mHz band. Longer windows and overlapping windows are natural extensions.

We use a single Michelson-like channel as a simplified setting. Standard LISA analyses use TDI combinations such as X,Y,Z or A,E,T. Using multiple TDI channels would provide additional information and could potentially improve absolute performance, however it would also introduce another experimental variable. Since the present work compares scoring rules, using a single channel isolates the methodological effect. Using TDI combinations will be explored in a follow up study because they provide multiple correlated observations of the same underlying signal population. From the perspective of representation learning, such redundancy is often advantageous because multi-channel autoencoders are capable of exploiting shared structures across channels while suppressing channel specific noise. This potentially will produce latent spaces with improved class separation and more informative density structure.

\subsection*{Instrumental noise model}

Instrumental noise is synthesized in the frequency domain from an analytic one-sided PSD with acceleration and optical metrology contributions, where $f$ denotes the gravitational-wave frequency. We define
\begin{equation}
S_{\mathrm{acc}}(f)=(3\times10^{-15})^2
\left(1+\left(\frac{4\times10^{-4}}{f}\right)^2\right)
\left(1+\left(\frac{f}{8\times10^{-3}}\right)^4\right),
\end{equation}
\begin{equation}
S_{\mathrm{oms}}(f)=(15\times10^{-12})^2,
\end{equation}
and combine them as
\begin{equation}
S_n(f)=\frac{10}{3L^2}
\left[
S_{\mathrm{oms}}(f)+
\frac{(3+\cos(2f/f_\ast))S_{\mathrm{acc}}(f)}
{(2\pi f)^2}
\right],
\end{equation}
where $L=2.5\times10^9\ \mathrm{m}$ is the LISA arm length and $f_\ast=c/(2\pi L)=19.09\ \mathrm{mHz}$ is the transfer frequency.

For a segment of duration \(T\), with frequency resolution \(\Delta f = 1/T\) and discrete Fourier frequencies \(f_k = k\Delta f\), we generate independent complex Gaussian samples
\begin{equation}
\tilde{n}(f_k)
=
(a_k+i b_k)
\sqrt{\frac{S_n(f_k)}{2\Delta f}},
\qquad
a_k,b_k\sim\mathcal{N}(0,1),
\end{equation}
where \(a_k\) and \(b_k\) are independent standard normal variates. The resulting frequency-domain samples are colored by the target PSD \(S_n(f)\) and transformed back to the time domain via an inverse real FFT to obtain the instrumental noise realization \(n(t)\). PSD values are interpolated onto the segment FFT frequency bins prior to coloring.

\subsection*{Waveform models and parameter draws}

Waveforms are generated analytically. For MBHB sources, we employ a
post-Newtonian-inspired chirp model. The instantaneous gravitational-wave
frequency evolves as

\begin{equation}
f(t)=f_{\mathrm{start}}
\left(1-\frac{|t-t_c|}{\tau}\right)^{-3/8},
\end{equation}
where \(f_{\mathrm{start}}\) is the initial frequency and \(t_c\) is the
coalescence time. The inspiral timescale is
\begin{equation}
\tau=
\frac{5c^5}
{256(\pi f_{\mathrm{start}})^{8/3}(GM_c)^{5/3}},
\end{equation}
with chirp mass
\begin{equation}
M_c=
\frac{(m_1m_2)^{3/5}}
{(m_1+m_2)^{1/5}},
\end{equation}
where \(m_1\) and \(m_2\) are the component masses. The waveform phase is
obtained by integrating the instantaneous frequency and projected onto
plus and cross polarizations with inclination-dependent amplitudes.

Source parameters are sampled uniformly with primary mass
\(m_1\in[10^4,10^7]\,M_\odot\), mass ratio
\(q=m_2/m_1\in[0.1,1]\), luminosity distance
\(d_L\in[1,20]\,\mathrm{Gpc}\), and initial frequency
\(f_{\mathrm{start}}\in[10^{-4},10^{-2}]\,\mathrm{Hz}\). The secondary
mass is determined by \(m_2=q\,m_1\).

For EMRI sources, we approximate the orbital motion as quasi-circular. The
orbital radius is parameterized as \(r=pr_g\), where
\(p\) is the dimensionless orbital radius and
\(r_g=GM/c^2\) is the gravitational radius of the central black hole of mass
\(M\). The orbital and dominant gravitational-wave frequencies are

\begin{equation}
f_{\mathrm{orb}}
=
\frac{1}{2\pi}
\sqrt{\frac{GM}{r^3}},
\qquad
f_{\mathrm{gw}}=2f_{\mathrm{orb}}.
\end{equation}
Frequency evolution is modeled with a leading-order radiation-reaction chirp,
\begin{equation}
\dot f
\propto
(\pi f_{\mathrm{gw}})^{11/3}
\left(\frac{GM}{c^3}\right)^{5/3}
\frac{\mu}{M},
\end{equation}
where \(\mu\) denotes the compact-object mass. The waveform is synthesized
as a finite sum of harmonics \(n=1,\ldots,5\) with amplitudes weighted by
the orbital eccentricity \(e\).

Source parameters are sampled uniformly with central black-hole mass
\(M\in[10^5,10^7]\,M_\odot\), compact-object mass
\(\mu\in[1,100]\,M_\odot\), dimensionless orbital radius
\(p\in[6,20]\), eccentricity
\(e\in[0,0.5]\), and luminosity distance
\(d_L\in[0.5,10]\,\mathrm{Gpc}\).

For galactic binaries, we employ a slowly evolving monochromatic model.
The instantaneous gravitational-wave frequency is parameterized by an
initial frequency \(f_0\) and a constant frequency derivative \(\dot f\),
giving
\begin{equation}
f(t)=f_0+\dot f\,t,
\qquad
\phi(t)=2\pi\left(f_0 t+\frac{1}{2}\dot f\,t^2\right)+\phi_0,
\end{equation}
where \(\phi_0\) is the initial phase. The waveform amplitude is denoted by
\(A\), and plus and cross polarizations are projected according to the
binary inclination \(\iota\). Source parameters are sampled uniformly with
initial frequency
\(f_0\in[10^{-4},10^{-2}]\,\mathrm{Hz}\),
dimensionless strain amplitude
\(A\in[10^{-22},10^{-20}]\),
frequency derivative
\(\dot f\in[-10^{-14},10^{-15}]\,\mathrm{Hz\,s^{-1}}\),
initial phase
\(\phi_0\in[0,2\pi)\), and isotropic inclination
\(\cos\iota\sim U[-1,1]\).

\subsection*{Confusion and resolvable-source construction}

Confusion foreground segments are defined as
\begin{equation}
x_{\mathrm{bg}}(t)=n(t)+\sum_{j=1}^{N_c} h^{(j)}_{\mathrm{GB}}(t),
\end{equation}
where \(n(t)\) is the instrumental noise realization and each
\(h^{(j)}_{\mathrm{GB}}(t)\) is a weak galactic-binary waveform rescaled to
a target SNR drawn uniformly from a confusion interval. In the main run,
we set \(N_c=1000\) and draw individual confusion-source SNRs from
\([0.1,2.0]\). Because the confusion foreground is the superposition of many
weak sources, its cumulative SNR can be substantially larger than the SNR
of any single constituent, ranging between approximately \([7,141]\).

SNR is computed against the same PSD by
\begin{equation}
\mathrm{SNR}^2 = 4\sum_k \frac{|\tilde h(f_k)|^2}{S_n(f_k)}\Delta f,
\end{equation}
and each waveform is amplitude-scaled by \(\mathrm{SNR}_{\mathrm{target}}/\mathrm{SNR}_{\mathrm{current}}\).

Positive test segments are built by adding one resolvable source:
\begin{equation}
x_{\mathrm{sig}}(t)=x_{\mathrm{bg}}(t)+h_{\mathrm{res}}(t),
\end{equation}
with \(\mathrm{SNR}_{\mathrm{target}}\sim\mathcal{U}(10,50)\). Labels are binary:
foreground-only (\(y=0\)) and foreground+resolvable (\(y=1\)).

\subsection*{Dataset composition and reproducibility}

For the benchmark configuration, we sample 5000 training confusion foreground segments, 200 test foreground segments, and 400 test signal segments. The 5000 foreground segments refer to the number of training examples presented to the autoencoder, not the total number of binaries present in the simulated foreground itself. The goal was to learn a stable representation of the confusion foreground distribution rather than exhaustively sample every possible realization over and over. Once the reconstruction loss and latent structure stabilize, adding large numbers of statistically similar foreground segments provides diminishing returns for the representation-learning stage. With more sophisticated simulations with more a more complex confusion foreground, the number of training foreground segments would need to increase accordingly. 

Signal-type fractions are MBHB 50\%, EMRI 30\%, and bright galactic binary 20\%. Seeded generation is deterministic at segment level (base seed plus index offsets), and per-segment metadata (source type, sampled parameters, and target SNR) are saved with the HDF5 datasets. This allows exact regeneration and audit of train/test composition.

\subsection*{CWT Representation}

Following Ref.~\cite{Cain2025}, one-hour segments are transformed into Morlet CWT scalograms spanning 0.1-100 mHz. Using 140 logarithmically spaced scales, the resulting magnitude scalograms are resized to 100x3600, log-transformed, and globally normalized. Full preprocessing details are inherited from Ref.~\cite{Cain2025}.

\section{Autoencoders and Geometry}
\label{sec:autoencoder}

\begin{figure}[htbp]
    \centering
    \includegraphics[width=0.50\textwidth]{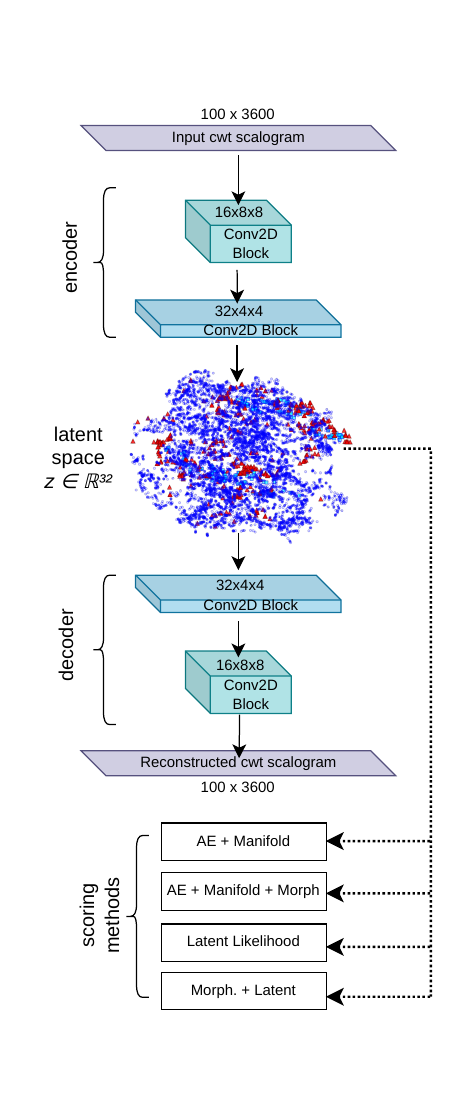}
    \caption[CNN autoencoder architecture with manifold learning]{
        Architecture of the CNN-based autoencoder with manifold learning for LISA
        gravitational-wave source separation. The encoder (top path) processes CWT
        scalograms ($100 \times 3600$) through two convolutional layers (Conv2d) with adaptive
        pooling, followed by linear layers to produce a 32-dimensional latent representation
        $z$ here shown as a 2D visualization. The decoder (bottom path) reconstructs the input through linear and
        transposed convolutional layers. Reconstruction error $\epsilon(x)$ is computed
        from the difference between input and reconstructed scalograms. In parallel, the
        manifold learning an likelihood-based scoring branch (bottom) operates on the latent space.During training, the k-NN index is built on latents from
        confusion foreground data only.
    }
    \label{fig:architecture}
\end{figure}

Following ref.~\cite{Cain2025}, we employ a lightweight convolutional autoencoder with two convolutional blocks and a 32-dimensional bottleneck tasked with learning a low-dimensional representation of LISA-like CWT scalograms. Approximately 33k trainable parameters are optimized with MSE reconstruction loss. A schematic is shown in Fig.~\ref{fig:architecture}. The autoencoder provides reconstruction error as an anomaly score, and it learns a latent space where we can analyze the structure of the data manifold. This geometric perspective, motivates our scoring approach.

From a differential geometry perspective, the autoencoder learns a chart from the data space to a low-dimensional latent space. Let $\mathcal{X} \subset \mathbb{R}^n$ denote the space of CWT scalograms, where $n = 100 \times 3600 = 360{,}000$ for our simulated LISA data. According to the manifold hypothesis, the training data (confusion foreground) lies near a smooth manifold $\mathcal{M} \subset \mathcal{X}$ of much lower intrinsic dimension $d \ll n$. The encoder $\phi:\mathcal{X}\rightarrow\mathcal{Z}$ maps each input segment to a latent vector
\[
z=\phi(x)\in\mathcal{Z},
\]
where the latent space is $\mathcal{Z}=\mathbb{R}^d$ with $d=32$ in our implementation. For data $x\in\mathcal{M}$, the encoder approximates a chart providing local coordinates on the manifold, while the decoder $\psi:\mathcal{Z}\rightarrow\mathcal{X}$ approximates the inverse chart, mapping latent vectors back to the data space. 

Throughout the remainder of this paper, the term \emph{local latent geometry} refers to the neighborhood relationships among latent vectors, while the \emph{global latent distribution} refers to the probability distribution of foreground latent vectors throughout the latent space. Likelihood-based methods estimate the associated latent density \(p(z)\).

The composite map $\psi \circ \phi: \mathcal{X} \to \mathcal{X}$ attempts to reconstruct points on the manifold, with reconstruction error $\epsilon(x) = \|x - \psi(\phi(x))\|^2$ measuring how well the point lies on the learned manifold. From a statistical perspective, the reconstruction error $\epsilon(x)$ acts as a test statistic for detecting deviations from the learned confusion-foreground manifold. Segments that lie close to the foreground manifold produce small reconstruction errors, while segments containing additional signal structure tend to yield larger values of this statistic. Standard detection metrics such as the ROC-AUC can then be used to evaluate the resulting trade-off between detection efficiency and false-alarm rate.

This geometric framework lends itself naturally to anomaly detection. In a vanilla autoencoder, points lying near the learned manifold (confusion foreground) are expected to exhibit low reconstruction error, while points lying farther from the manifold (resolvable sources) should produce larger errors. However, reconstruction error alone provides only a weak notion of manifold proximity and does not explicitly exploit local geometric structure. In \cite{Cain2025}, we found that off-manifold geometric scores based on latent-space neighborhoods improved ROC-AUC by approximately 35\% relative to reconstruction error, demonstrating that explicit local latent geometry contains information discarded by the reconstruction objective. Nevertheless, these methods remain fundamentally local. Although they capture distances relative to nearby regions of the manifold, they do not account for the global arrangement and distribution of latent neighborhoods. This observation motivates scoring directly on the latent representation itself. We therefore compare local geometric scores with density-based scores that exploit global latent distribution, as discussed in Section~\ref{sec:methods}.

\section{One-class scoring rules}
\label{sec:methods}

Here we explore the one-class scoring rules in Table~\ref{tab:method_summary}. The purpose of this section is to separate representation from decision rule: all methods consume the same preprocessed segments, but they define anomaly score in different ways. The task is straightforward: in a confusion-limited setting, which one-class scoring rule gives the best ranking of foreground-only versus foreground-plus-resolvable segments?

\begin{table}[htbp]
\centering
\caption{Method summary with score definitions under the fixed protocol.}
\label{tab:method_summary}
\begin{tabular}{p{3.1cm}p{3.9cm}p{2.6cm}}
\toprule
Method & Score definition & One-class model \\
\midrule
AE+Manifold & \(s_{\mathrm{base}}=\alpha e_{\mathrm{AE}}+\beta d_{\mathcal{M}}\) & Local latent geometry \\
AE+Manifold+Morph & \(s_{\mathrm{fuse}}=(1-\lambda)s_{\mathrm{base}}+\lambda d_{\phi}\) & Geometry + morphology kNN \\
Likelihood (Latent) & \(s_{\mathrm{lik}}=-\log p(f_\theta(x))\) & GMM or KDE on latent \\
Likelihood (Morph+Latent) & \(s=-\log p([\phi(x),f_\theta(x)])\) & KDE on fused feature vector \\
\bottomrule
\end{tabular}
\end{table}

\subsection*{Experimental protocol}

All method comparisons are presented under identical conditions.  By identical we mean that the same latent space representation of the confusion foreground was presented to each method and identical test data was scored by each method.  The only difference in each workflow was scoring as shown in Fig.~\ref{fig:architecture}. To assess robustness to training stochasticity, we repeat each reportable method across three independent training seeds. Final performance is reported on the same held-out test set per seed for all methods.

Primary comparisons are based on ranking metrics computed from full test-score distributions.  
For seed-robustness reporting, we present aggregate statistics across the three runs (mean and standard deviation), alongside per-seed values in supplementary material.

\subsection*{Baseline Geometry Framework}
This baseline geometry formulation and its confusion-limited motivation were introduced in our prior study \cite{Cain2025}. In this study, the optimal combination of \(\alpha\) and \(\beta\), defined below, yielded a 35\% improvement in ROC-AUC over autoencoder-only detection. The baseline model is an autoencoder trained only on foreground segments. 

Let \(x\) denote a CWT input and \(z=f_\theta(x)\) its latent vector. The decoder \(g_\theta\) reconstructs \(\hat{x}=g_\theta(z)\), and reconstruction error is
\begin{equation}
e_{\mathrm{AE}}(x)=\|x-\hat{x}\|^2.
\end{equation}

A latent manifold is then estimated from foreground latents using a local \(k\)-nearest-neighbor construction.  
For a test latent \(z\), we compute a local off-manifold deviation \(d_{\mathcal{M}}(z)\), interpreted as distance from typical foreground geometry.

The baseline anomaly score is
\begin{equation}
s_{\mathrm{base}}(x)=\alpha\,e_{\mathrm{AE}}(x)+\beta\,d_{\mathcal{M}}(f_\theta(x)),
\end{equation}
with \(\alpha,\beta\ge 0\).  
Operationally, \(e_{\mathrm{AE}}\) captures reconstruction mismatch, while \(d_{\mathcal{M}}\) measures deviation from the local latent geometry relative to foreground structure.

\subsection*{Physics-Feature Augmentation}

We next add morphology descriptors computed from the same CWT inputs.  
These features summarize physically relevant time-frequency structure (track continuity, concentration, anisotropy, and related shape cues) that can be weakly expressed in raw latent distance alone.

From foreground morphology vectors, we build a one-class reference set. Let
\(\phi_j\) denote the morphology vector of the \(j\)-th foreground reference
sample. Each test sample is then scored by its average distance to the
\(k\) nearest foreground reference vectors in morphology space:
\begin{equation}
d_{\phi}(x)=\frac{1}{k}\sum_{j=1}^{k}\|\phi(x)-\phi_j\|_2.
\end{equation}

We refer to this weighted combination of geometric and morphology scores as fusion because it combines complementary information from latent geometry and hand-crafted morphology descriptors. A fused score is then defined by
\begin{equation}
s_{\mathrm{fuse}}(x)=(1-\lambda)\,s_{\mathrm{base}}(x)+\lambda\,d_{\phi}(x),\qquad \lambda\in[0,1].
\end{equation}
This tests whether explicit morphology contributes information beyond latent geometry and reconstruction alone.

\subsection*{Likelihood One-Class Scoring in Latent Space}

The final method replaces distance-to-manifold scoring with explicit density modeling in latent space.  
Using foreground latents \(z\sim p_{\mathrm{bg}}(z)\), we fit one-class likelihood models and score by negative log-likelihood.

For Gaussian mixtures,
\begin{equation}
p_{\mathrm{GMM}}(z)=\sum_{m=1}^{k}\pi_m\,\mathcal{N}(z\mid \mu_m,\Sigma_m),
\end{equation}
where \(k\) is the number of mixture components, \(\pi_m\) is the mixture
weight of component \(m\), and \(\mathcal{N}(z\mid \mu_m,\Sigma_m)\) is a
Gaussian density with mean \(\mu_m\) and covariance matrix \(\Sigma_m\).
The parameters \(\{\pi_m,\mu_m,\Sigma_m\}_{m=1}^{k}\) are estimated using
expectation-maximization \cite{Dempster1977}.

For kernel density estimation,
\begin{equation}
p_{\mathrm{KDE}}(z)=\frac{1}{Nh^d}\sum_{i=1}^{N}K\!\left(\frac{z-z_i}{h}\right),
\end{equation}
where \(N\) is the number of foreground latent vectors,
\(z_i\) is the \(i\)-th foreground latent vector,
\(K(\cdot)\) is the kernel function,
\(h\) is the bandwidth, and
\(d\) is the latent-space dimension.
Following the nonparametric estimator in \cite{Parzen1962}, kernel density estimation avoids assuming a finite mixture model. Instead, each foreground latent vector contributes a smooth kernel centered at its location. The bandwidth parameter \(h\) determines the degree of smoothing: small values emphasize local structure, whereas large values recover a smoother global latent density estimate.

The anomaly score is
\begin{equation}
s_{\mathrm{lik}}(x)=-\log p(f_\theta(x)),
\end{equation}
where \(p\) is either \(p_{\mathrm{GMM}}\) or \(p_{\mathrm{KDE}}\).  
Conceptually, likelihood scoring measures how probable a latent vector is under the entire foreground distribution. This distinction between local latent geometry and global density is central to this work.

\section{Results}
\label{sec:results}
\subsection*{Performance Metrics}
The goal of this study is to compare one-class scoring rules. With that in mind, we emphasize threshold-independent ranking metrics, namely ROC-AUC and PR-AUC. These metrics evaluate how effectively a scoring function separates foreground-only segments from foreground plus resolvable segments across all possible decision thresholds. We avoid a fixed threshold to allow for direct comparison of different scoring rules without conflating ranking quality with threshold calibration.

This is different from the standard gravitational-wave search practice, where significance is assessed at a fixed false-alarm probability or false-alarm rate and performance is reported via detection efficiency or sensitivity at a chosen operating point. Those metrics are appropriate for deployed search pipelines, but they depend on threshold selection and foreground estimation. Since the purpose of this work is to isolate the effect of the scoring rule under identical conditions, threshold-free metrics provide a more appropriate comparison. Something closer to standard gravitational wave detection practice would be to  calibrate thresholds using foreground-only validation data and report detection efficiency at fixed false-positive rates or significance levels. 

We also report precision and recall at a single operating point. The threshold is taken to be the median anomaly score on the evaluation set, applied independently for each method and seed,
\[
\hat{y}=\mathbf{1}\left\{s(x)>\mathrm{median}(s)\right\}.
\]
This rule is identical across all methods and should be viewed only as a protocol-dependent diagnostic. Consequently, precision and recall are secondary quantities, while ROC-AUC and PR-AUC remain the principal metrics used to compare scoring performance.
\subsection*{Primary Results and Interpretation}

\begin{figure*}[htbp]
\centering
\begin{minipage}[t]{0.49\textwidth}
\centering
\includegraphics[width=\textwidth]{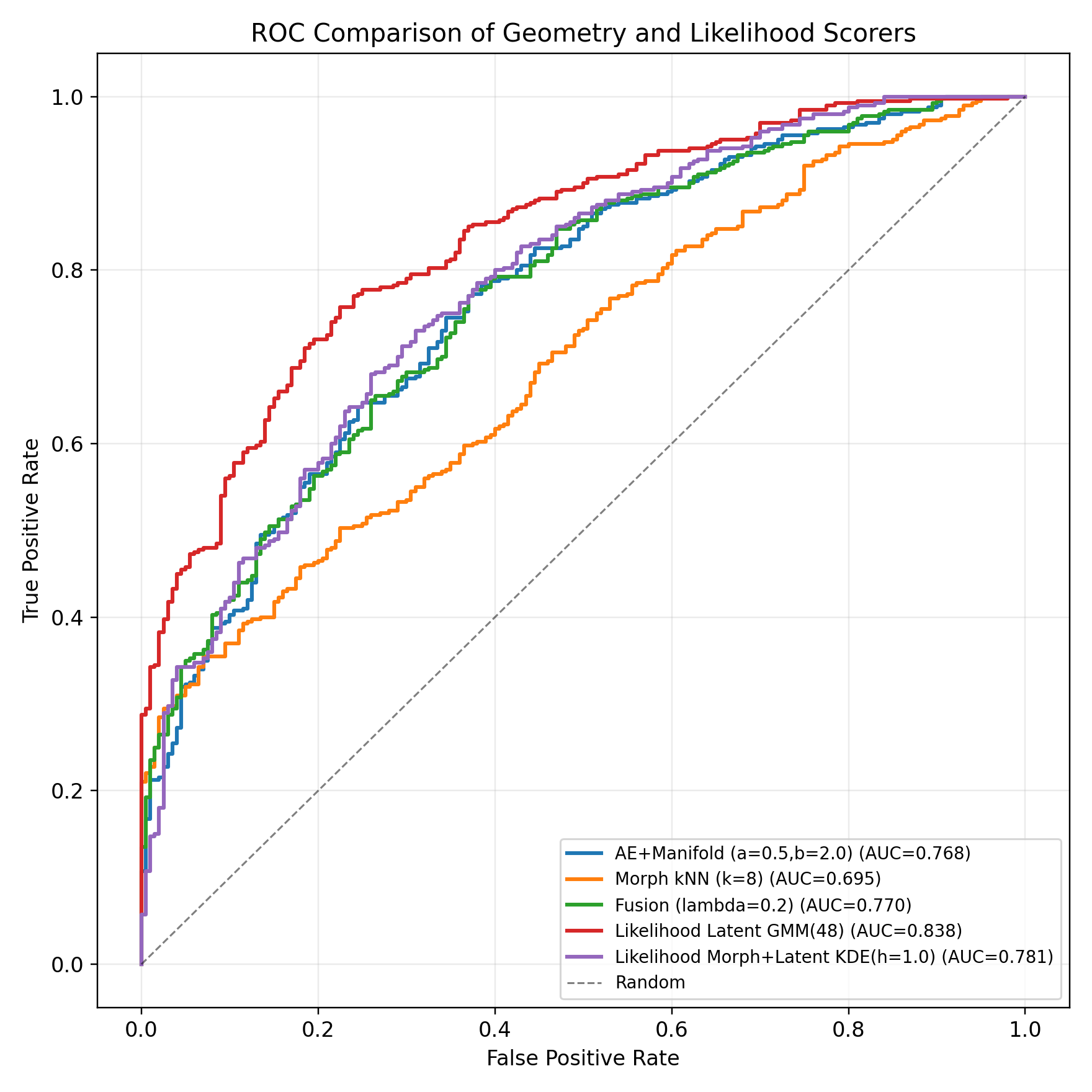}
\end{minipage}
\hfill
\begin{minipage}[t]{0.49\textwidth}
\centering
\includegraphics[width=\textwidth]{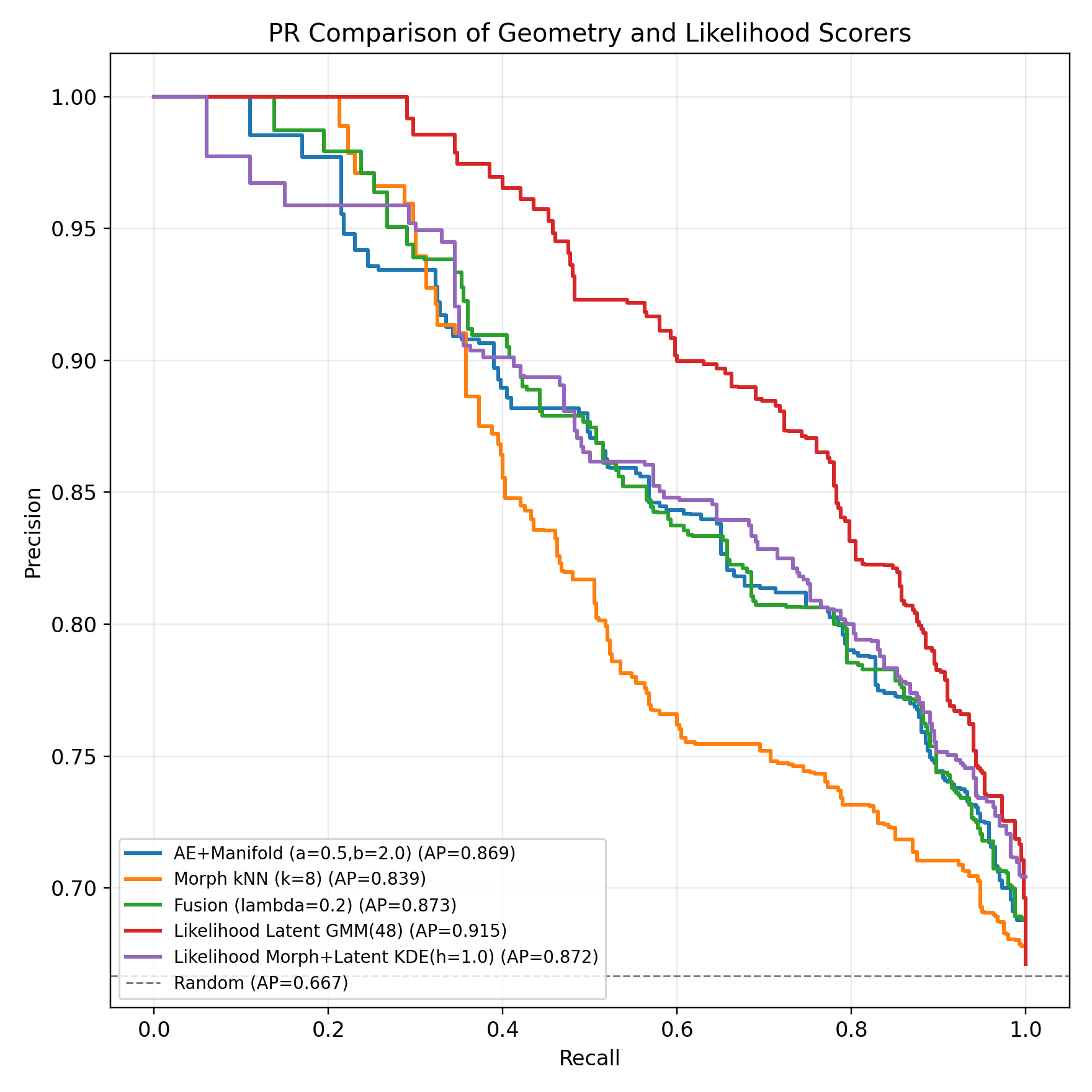}
\end{minipage}
\caption{
Representative single-run ranking curves across methods.
Left: ROC (TPR vs FPR). Right: precision-recall.
Legend values correspond to threshold-free summary scores for that run.
}
\label{fig:prelim_perf_curves}
\end{figure*}

Figure~\ref{fig:prelim_perf_curves} shows representative single-run ROC and PR curves for the main methods. Table~\ref{tab:main_results} reports three-seed aggregates (mean $\pm$ std) for the primary comparisons, with morphology-only included as a diagnostic baseline.

Several observations emerge immediately. First, likelihood-based scoring in latent space provides the strongest overall performance. Relative to the AE+Manifold baseline, latent likelihood improves the mean ROC-AUC from $0.7663$ to $0.8555$ and the mean PR-AUC from $0.8667$ to $0.9219$. These gains are substantially larger than the corresponding seed-to-seed variation, suggesting that the improvement is systematic rather than a consequence of training stochasticity.

Second, augmenting the geometry baseline with morphology descriptors produces only modest gains. The AE+Manifold+Morph method improves ROC-AUC and PR-AUC slightly relative to AE+Manifold, but the effect size is small. This suggests that morphology features provide complementary information, although they do not fundamentally alter the ranking behavior.

Third, combining morphology with likelihood does not improve over latent likelihood alone. The Morph+Latent likelihood variant performs better than the geometry baseline but remains below the latent-only likelihood model. This behavior suggests that much of the discriminative information contained in the hand-crafted morphology descriptors is already encoded within the learned latent representation. Adding explicit morphology features therefore introduces little additional benefit.

The morphology-only kNN baseline performs worst among all methods. While morphology descriptors contain useful information, their performance indicates that they are insufficient as a stand-alone representation. Learned latent representations therefore appear to capture structure beyond what is available through hand-crafted time-frequency descriptors alone.

Finally, the likelihood-based model exhibits lower seed-to-seed variability than the geometry baseline. In addition to achieving higher average ROC-AUC and PR-AUC, the latent likelihood model produces smaller standard deviations across seeds. This suggests that global density modeling is not only more accurate in this benchmark but also more stable. Taken together, the results support the central conclusion of this work: source-resolvability information in learned latent representations is organized globally, and exploiting the global latent distribution yields stronger discrimination than relying on local geometric deviation alone.

\begin{table*}[htbp]
\centering
\caption{
Primary performance comparison under fixed-config reporting.
Values are mean $\pm$ std over seeds 101/202/303.
PR-AUC is reported as average precision (AP).
Precision and recall use the fixed median-score decision rule.
Fixed settings: baseline/fusion use $\alpha=0.5,\beta=2.0$; fusion uses $\lambda=0.2$; latent likelihood uses GMM(48); morph+latent likelihood uses KDE($h=1.0$); morph-only uses kNN($k=8$).
}
\label{tab:main_results}
\renewcommand{\arraystretch}{1.1}
\setlength{\tabcolsep}{3pt}
\begin{tabular}{p{4.2cm}cccc}
\toprule
Method & ROC-AUC & PR-AUC (AP) & Precision & Recall \\
\midrule
AE+Manifold & $0.7663 \pm 0.0450$ & $0.8667 \pm 0.0255$ & $0.8411 \pm 0.0269$ & $0.6308 \pm 0.0202$ \\
AE+Manifold+Morph & $0.7688 \pm 0.0419$ & $0.8716 \pm 0.0223$ & $0.8389 \pm 0.0252$ & $0.6292 \pm 0.0189$ \\
\textbf{Likelihood GMM (Latent)} & $\mathbf{0.8555 \pm 0.0181}$ & $\mathbf{0.9219 \pm 0.0118}$ & $\mathbf{0.8967 \pm 0.0167}$ & $\mathbf{0.6725 \pm 0.0125}$ \\
Likelihood KDE (Latent) & $0.8033 \pm 0.0159$ & $0.8836 \pm 0.0109$ & $0.8489 \pm 0.0107$ & $0.6367 \pm 0.0080$ \\
Morph kNN & $0.6947 \pm 0.0000$ & $0.8388 \pm 0.0000$ & $0.7667 \pm 0.0000$ & $0.5750 \pm 0.0000$ \\
\bottomrule
\end{tabular}
\renewcommand{\arraystretch}{1.0}
\end{table*}

\subsection*{Likelihood Hyperparameter Ablation}

\begin{figure}[htbp]
\centering
\includegraphics[width=0.95\linewidth]{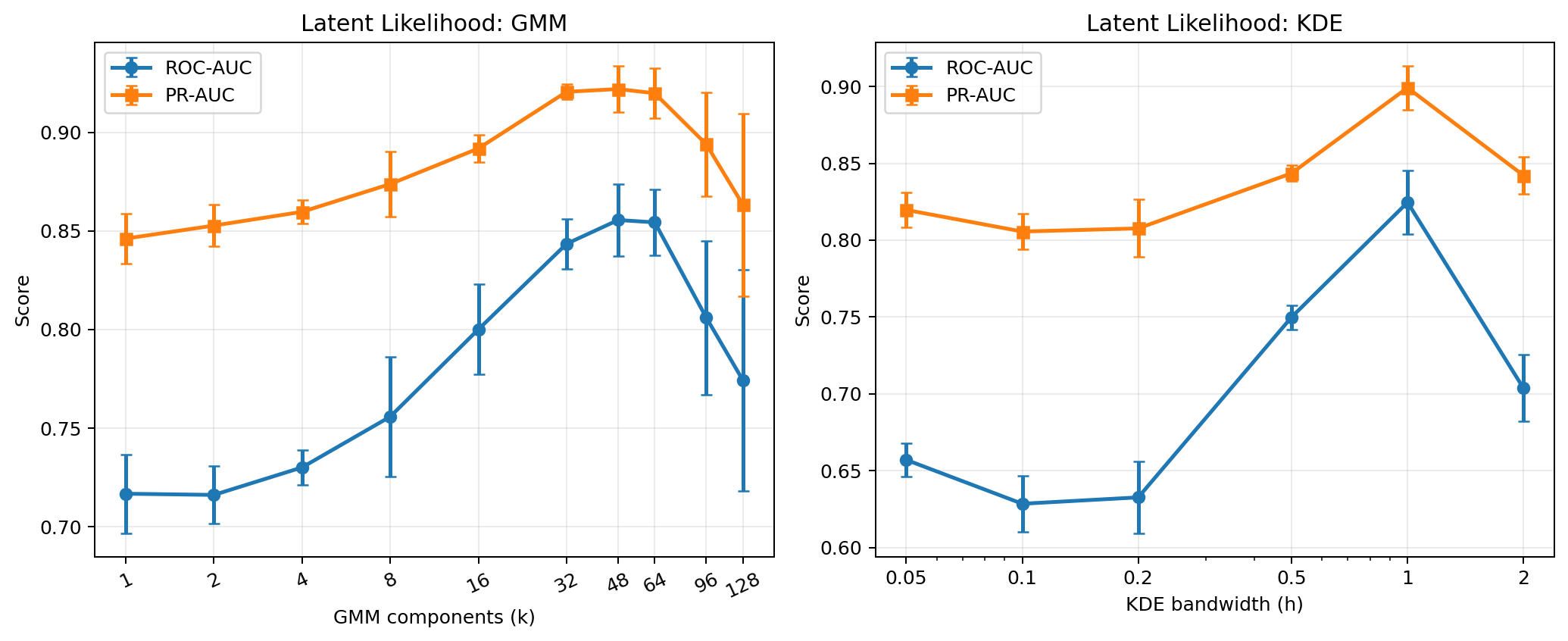}
\caption{Latent-likelihood ablation. Left: ROC-AUC and PR-AUC versus GMM component count. Right: ROC-AUC and PR-AUC versus KDE bandwidth (log-scaled). Values are mean $\pm$ std across three seeds.}
\label{fig:likelihood_ablation}
\end{figure}

Table~\ref{tab:likelihood_ablation} summarizes the effect of model complexity on latent-likelihood performance, and Figure~\ref{fig:likelihood_ablation} provides a visual summary. Two trends are immediately apparent.

For Gaussian mixture models, performance improves steadily as the number of mixture components increases. Moving from a single Gaussian to progressively richer mixture models produces consistent gains in both ROC-AUC and PR-AUC, reaching a maximum at $M=48$ components. Beyond this point, performance degrades, with larger models exhibiting increased variance across seeds. This behavior is consistent with a bias-variance tradeoff: too few components underfit the global latent distribution, while excessive complexity begins to overfit finite training samples.

Importantly, the monotonic improvement observed between $k=1$ and $k=48$ suggests that the global latent distribution is not well described by a single Gaussian. Instead, the results support the interpretation that the latent representation possesses meaningful multi-modal global latent distribution that is better captured by increasingly expressive density models. If local latent geometry alone contained all the relevant information, one would not expect progressively richer global density models to continue improving performance. The fact that increasing GMM complexity improves discrimination up to \(k= 48\) suggests that source-resolvability information is distributed across the global latent distribution rather than solely by local latent geometry.

Kernel density estimation exhibits a similar tradeoff. Small bandwidths ($h \le 0.2$) produce poor ranking performance, indicating that overly local density estimates are sensitive to noise and fail to capture the broader organization of latent space. Increasing the bandwidth improves both ROC-AUC and PR-AUC, with optimal performance obtained at $h=1.0$. Further smoothing ($h=2.0$) reduces performance, suggesting that excessive smoothing obscures important structure in the latent distribution.

Although KDE provides competitive performance, its sensitivity to bandwidth selection is substantially greater than that of the GMM approach. In contrast, GMM performance remains relatively stable near the optimum, with little difference between $k=32$, $k=48$, and $k=64$. Consequently, the GMM model appears to provide a more robust parameterization of the global latent distribution for the present benchmark.

Taken together, these ablation results provide additional support for the central thesis of this work. The improvement obtained with increasingly expressive density models suggests that useful source-resolvability information is encoded in the global latent distribution rather than solely in local latent relationships.

\begin{table}[htbp]
\centering
\caption{Latent-likelihood ablation on hyperparameters (mean $\pm$ std over 3 seeds).}
\label{tab:likelihood_ablation}
\renewcommand{\arraystretch}{1.05}
\setlength{\tabcolsep}{4pt}
\small
\begin{tabular}{lcc}
\toprule
Setting & ROC-AUC & PR-AUC (AP) \\
\midrule
\multicolumn{3}{l}{\textit{GMM components}} \\
$k=1$   & $0.7168 \pm 0.0200$ & $0.8461 \pm 0.0127$ \\
$k=2$   & $0.7162 \pm 0.0145$ & $0.8527 \pm 0.0106$ \\
$k=4$   & $0.7302 \pm 0.0087$ & $0.8597 \pm 0.0058$ \\
$k=8$   & $0.7559 \pm 0.0304$ & $0.8738 \pm 0.0166$ \\
$k=16$  & $0.8001 \pm 0.0228$ & $0.8918 \pm 0.0070$ \\
$k=32$  & $0.8435 \pm 0.0127$ & $0.9205 \pm 0.0039$ \\
\textbf{$k=48$}  & $\mathbf{0.8555 \pm 0.0181}$ & $\mathbf{0.9219 \pm 0.0118}$ \\
$k=64$  & $0.8543 \pm 0.0167$ & $0.9197 \pm 0.0128$ \\
$k=96$  & $0.8060 \pm 0.0390$ & $0.8939 \pm 0.0265$ \\
$k=128$ & $0.7743 \pm 0.0560$ & $0.8632 \pm 0.0462$ \\
\midrule
\multicolumn{3}{l}{\textit{KDE bandwidth}} \\
$h=0.05$ & $0.6571 \pm 0.0109$ & $0.8197 \pm 0.0112$ \\
$h=0.1$  & $0.6286 \pm 0.0183$ & $0.8056 \pm 0.0115$ \\
$h=0.2$  & $0.6327 \pm 0.0233$ & $0.8076 \pm 0.0187$ \\
$h=0.5$  & $0.7499 \pm 0.0079$ & $0.8435 \pm 0.0053$ \\
\textbf{$h=1.0$} & $\mathbf{0.8246 \pm 0.0208}$ & $\mathbf{0.8991 \pm 0.0143}$ \\
$h=2.0$  & $0.7037 \pm 0.0217$ & $0.8421 \pm 0.0123$ \\
\bottomrule
\end{tabular}
\normalsize
\renewcommand{\arraystretch}{1.0}
\end{table}

\subsection*{Robustness and Statistical Validation}

\begin{table}[htbp]
\centering
\caption{
Paired seed-wise metric deltas relative to the AE+Manifold baseline
($\alpha=0.5,\beta=2.0$) under fixed-config reporting.
Values are mean \(\pm\) std across seeds 101/202/303.
``Wins (AUC)'' counts seeds where the method's ROC-AUC exceeds baseline.
}
\label{tab:paired_deltas}
\begin{tabular}{lccc}
\toprule
Comparison & $\Delta$ROC-AUC & $\Delta$PR-AUC & Wins (AUC) \\
\midrule
Fusion ($\lambda=0.2$) & $+0.0025 \pm 0.0031$ & $+0.0050 \pm 0.0033$ & $3/3$ \\
Likelihood (Latent, GMM(48)) & $\mathbf{+0.0892 \pm 0.0293}$ & $\mathbf{+0.0552 \pm 0.0145}$ & $\mathbf{3/3}$ \\
Likelihood (Morph+Latent, KDE($h=1.0$)) & $+0.0370 \pm 0.0312$ & $+0.0169 \pm 0.0176$ & $3/3$ \\
\bottomrule
\end{tabular}
\end{table}

Table~\ref{tab:paired_deltas} summarizes paired seed-wise improvements relative to the AE+Manifold baseline. Because all methods share identical latent space and test data and differ only in their scoring rules, these paired comparisons isolate the effect of each scoring rule.

Latent likelihood scoring performs consistently higher. Using a fixed GMM(48) configuration, latent likelihood improves upon the AE+Manifold baseline in all seeds, producing mean gains of $+0.0892$ ROC-AUC and $+0.0552$ PR-AUC. Importantly, these gains are substantially larger than the seed-to-seed variation reported in Table~\ref{tab:main_results}, which suggests that the improvement is systematic rather than due to training stochasticity. 

In contrast, morphology fusion produces only modest but directionally consistent improvements. At fixed $\lambda=0.2$, fusion exceeds the baseline in all three seeds, although the mean gains stay small ($+0.0025$ ROC-AUC and $+0.0050$ PR-AUC). This suggests that morphology provides complementary information but with smaller effect size than latent density modeling.

The morphology+latent likelihood variant also consistently outperforms the geometry baseline, with mean gains of $+0.0370$ ROC-AUC and $+0.0169$ PR-AUC. However, its aggregate performance remains below latent-only likelihood. This is consistent with the interpretation that much of the useful morphology information is already encoded in the learned latent representation, which leaves limited additional information to be gained from the morphology features.

The morphology-only kNN model is deterministic under identical conditions and is only included as a diagnostic reference. Repeated seed entries are identical by construction.

Figure~\ref{fig:prelim_perf_curves} presents representative single-run ROC and PR curves, though the principal conclusions of this work are based on the three-seed aggregates reported in Tables~\ref{tab:main_results} and~\ref{tab:paired_deltas}. Taken together, these results indicate that the superiority of latent likelihood scoring is not an isolated outcome from a favorable seed, but a reproducible property of the learned representation. This robustness further supports the central conclusion of the paper: source-resolvability information is encoded in global latent distribution and can be exploited more effectively through density modeling than through local geometric deviation alone.

\section{Conclusion and Future Work}

This paper compared one-class scoring strategies for resolvable-source ranking in confusion-limited synthetic LISA data under identical conditions. The strongest performer was likelihood-based scoring in latent space, with consistent gains over the AE+manifold baseline across all three seeds. Morphology fusion improved baseline performance, but with smaller effect size.

The main result shows that within a fixed CWT latent representation, global latent density modeling captures resolvability structure more effectively than local manifold-distance scoring. This supports a broader program of developing representation-aware, unsupervised machine-learning methods for gravitational-wave astronomy.

Several research directions follow from this work. First, the robustness of the observed likelihood advantage should be tested under a controlled distribution shift. Important axes are confusion level, source-population mixture, and resolvable-source SNR range. This will determine if its advantage persists under changes in the underlying data distribution.

Second, future studies should investigate richer representations using multi-channel TDI observables. Standard LISA analyses employ correlated channels such as $(A,E,T)$ or $(X,Y,Z)$, which provide multiple different views of the same underlying signal population. From a representation learning perspective, this redundancy could increase performance because multi-channel autoencoders can exploit shared structure while suppressing noise that is channel specific. This might lead to latent spaces with improved class separation and a more informative density structure.

Third, the global latent distribution identified in this work raises the question of coordinate dependence. The geometry of a learned latent space is not unique, and different linear transformations can alter Euclidean distances while preserving information content. An important next step is to investigate whether the observed density structure is invariant under rotations, whitening transformations, and other changes of coordinates. A study like this would help determine whether the likelihood advantage reflects intrinsic properties of the data or artifacts of a particular parameterization.

A related question is whether an intrinsic coordinate system exists for gravitational-wave representations. In traditional matched-filter analyses, waveform mismatch induces a local information geometry through the Fisher metric. A promising direction is to investigate whether learned latent representations can recover coordinate systems whose geometry aligns with mismatch-based notions of distance. Establishing such a connection would provide a bridge between representation learning and the information geometry underlying conventional gravitational-wave searches.

Finally, if latent likelihood continues to demonstrate consistent advantages after distribution shift and coordinate transformations, more sophisticated density models should be explored. More complex models should repeat the same discipline used throughout this study: reproducible, seed-robust improvements under identical conditions rather than isolated gains from single runs. The present work represents one step toward a broader framework in which representation learning, latent geometry, and information-theoretic structure might be integrated into future gravitational-wave data-analysis methods.

\section*{Code and Data Availability}

To ensure reproducibility, all experiments reported in this paper are tied to the public repository \url{https://github.com/jericho-cain/lisa-cwt-oneclass-likelihood} at tag \texttt{v1.0.0-paper}; the repository includes fixed seed configurations, evaluation scripts, and the figure/table summary artifacts used in the manuscript.

\bibliographystyle{iopart-num}
\bibliography{refs}

@Article{Chatterji2004,
  author    = {Chatterji, S and Blackburn, L and Martin, G and Katsavounidis, E},
  journal   = {Classical and Quantum Gravity},
  title     = {Multiresolution techniques for the detection of gravitational-wave bursts},
  year      = {2004},
  issn      = {1361-6382},
  month     = sep,
  number    = {20},
  pages     = {S1809--S1818},
  volume    = {21},
  doi       = {10.1088/0264-9381/21/20/024},
  publisher = {IOP Publishing},
}

@Article{Gabbard2018,
  author    = {Gabbard, Hunter and Williams, Michael and Hayes, Fergus and Messenger, Chris},
  journal   = {Physical Review Letters},
  title     = {Matching Matched Filtering with Deep Networks for Gravitational-Wave Astronomy},
  year      = {2018},
  issn      = {1079-7114},
  month     = apr,
  number    = {14},
  pages     = {141103},
  volume    = {120},
  doi       = {10.1103/physrevlett.120.141103},
  publisher = {American Physical Society (APS)},
}

@Article{George2018,
  author    = {George, Daniel and Huerta, E.A.},
  journal   = {Physics Letters B},
  title     = {Deep Learning for real-time gravitational wave detection and parameter estimation: Results with Advanced LIGO data},
  year      = {2018},
  issn      = {0370-2693},
  month     = mar,
  pages     = {64--70},
  volume    = {778},
  doi       = {10.1016/j.physletb.2017.12.053},
  publisher = {Elsevier BV},
}

@Article{Moreno2022,
  author    = {Moreno, Eric A and Borzyszkowski, Bartlomiej and Pierini, Maurizio and Vlimant, Jean-Roch and Spiropulu, Maria},
  journal   = {Machine Learning: Science and Technology},
  title     = {Source-agnostic gravitational-wave detection with recurrent autoencoders},
  year      = {2022},
  issn      = {2632-2153},
  month     = apr,
  number    = {2},
  pages     = {025001},
  volume    = {3},
  doi       = {10.1088/2632-2153/ac5435},
  publisher = {IOP Publishing},
}

@Article{Owen1996,
  author    = {Owen, Benjamin J.},
  journal   = {Physical Review D},
  title     = {Search templates for gravitational waves from inspiraling binaries: Choice of template spacing},
  year      = {1996},
  issn      = {1089-4918},
  month     = jun,
  number    = {12},
  pages     = {6749--6761},
  volume    = {53},
  doi       = {10.1103/physrevd.53.6749},
  publisher = {American Physical Society (APS)},
}

@Article{Powell2015,
  author    = {Powell, Jade and Trifirò, Daniele and Cuoco, Elena and Heng, Ik Siong and Cavaglià, Marco},
  journal   = {Classical and Quantum Gravity},
  title     = {Classification methods for noise transients in advanced gravitational-wave detectors},
  year      = {2015},
  issn      = {1361-6382},
  month     = oct,
  number    = {21},
  pages     = {215012},
  volume    = {32},
  doi       = {10.1088/0264-9381/32/21/215012},
  publisher = {IOP Publishing},
}

@Misc{Fayad2024,
  author        = {Fayad, Ammar},
  month         = nov,
  title         = {Unsupervised Learning Approach to Anomaly Detection in Gravitational Wave Data},
  year          = {2024},
  archiveprefix = {arXiv},
  copyright     = {Creative Commons Attribution 4.0 International},
  doi           = {10.48550/ARXIV.2411.19450},
  eprint        = {2411.19450},
  primaryclass  = {gr-qc},
  publisher     = {arXiv},
}

@Article{Klimenko2008,
  author    = {Klimenko, S and Yakushin, I and Mercer, A and Mitselmakher, G},
  journal   = {Classical and Quantum Gravity},
  title     = {A coherent method for detection of gravitational wave bursts},
  year      = {2008},
  issn      = {1361-6382},
  month     = may,
  number    = {11},
  pages     = {114029},
  volume    = {25},
  doi       = {10.1088/0264-9381/25/11/114029},
  publisher = {IOP Publishing},
}

@Misc{AmaroSeoane2017,
  author        = {{LISA Collaboration}},
  month         = feb,
  title         = {Laser Interferometer Space Antenna},
  year          = {2017},
  archiveprefix = {arXiv},
  copyright     = {arXiv.org perpetual, non-exclusive license},
  doi           = {10.48550/ARXIV.1702.00786},
  eprint        = {1702.00786},
  primaryclass  = {astro-ph.IM},
  publisher     = {arXiv},
}

@Article{Cornish2017,
  author        = {Cornish, Neil and Robson, Travis},
  journal       = {Journal of Physics: Conference Series},
  title         = {Galactic binary science with the new LISA design},
  year          = {2017},
  issn          = {1742-6596},
  month         = may,
  pages         = {012024},
  volume        = {840},
  archiveprefix = {arXiv},
  copyright     = {arXiv.org perpetual, non-exclusive license},
  date          = {2017-03-29},
  doi           = {10.1088/1742-6596/840/1/012024},
  eprint        = {1703.09858},
  primaryclass  = {astro-ph.IM},
  publisher     = {IOP Publishing},
}

@Article{Robson2019,
  author    = {Robson, Travis and Cornish, Neil J and Liu, Chang},
  journal   = {Classical and Quantum Gravity},
  title     = {The construction and use of LISA sensitivity curves},
  year      = {2019},
  issn      = {1361-6382},
  month     = apr,
  number    = {10},
  pages     = {105011},
  volume    = {36},
  doi       = {10.1088/1361-6382/ab1101},
  publisher = {IOP Publishing},
}

@Article{Timpano2006,
  author    = {Timpano, Seth E. and Rubbo, Louis J. and Cornish, Neil J.},
  journal   = {Physical Review D},
  title     = {Characterizing the galactic gravitational wave background with LISA},
  year      = {2006},
  issn      = {1550-2368},
  month     = jun,
  number    = {12},
  pages     = {122001},
  volume    = {73},
  doi       = {10.1103/physrevd.73.122001},
  publisher = {American Physical Society (APS)},
}

@Article{Nelemans2001,
  author    = {Nelemans, G. and Yungelson, L. R. and Portegies Zwart, S. F.},
  journal   = {Astronomy \& Astrophysics},
  title     = {The gravitational wave signal from the Galactic disk population of binaries containing two compact objects},
  year      = {2001},
  issn      = {1432-0746},
  month     = sep,
  number    = {3},
  pages     = {890--898},
  volume    = {375},
  doi       = {10.1051/0004-6361:20010683},
  publisher = {EDP Sciences},
}

@Article{Crowder2007,
  author    = {Crowder, Jeff and Cornish, Neil J.},
  journal   = {Physical Review D},
  title     = {Solution to the galactic foreground problem for LISA},
  year      = {2007},
  issn      = {1550-2368},
  month     = feb,
  number    = {4},
  pages     = {043008},
  volume    = {75},
  doi       = {10.1103/physrevd.75.043008},
  publisher = {American Physical Society (APS)},
}

@techreport{LISASciRD,
  author      = {{LISA Consortium}},
  title       = {LISA Science Requirements Document},
  institution = {European Space Agency},
  year        = {2018},
  number      = {ESA-L3-EST-SCI-RS-001},
}

@article{Cain2025,
  title = {Manifold Learning for Source Separation in Confusion-Limited Gravitational-Wave Data},
  author = {Cain, Jericho},
  year = 2026,
  month = jul,
  journal = {Classical and Quantum Gravity},
  volume = {43},
  number = {13},
  pages = {135031},
  publisher = {IOP Publishing},
  doi = {10.1088/1361-6382/ae83ff}
}

@article{Cain2026,
  title = {Template-Free Gravitational Wave Detection with {{CWT-LSTM}} Autoencoders: A Case Study of Run-Dependent Calibration Effects in {{LIGO}} Data},
  author = {Cain, Jericho},
  year = 2026,
  month = feb,
  journal = {Classical and Quantum Gravity},
  volume = {43},
  number = {3},
  pages = {035019},
  publisher = {IOP Publishing},
  doi = {10.1088/1361-6382/ae415e}
}

@Article{Dempster1977,
  author    = {Dempster, A. P. and Laird, N. M. and Rubin, D. B.},
  journal   = {Journal of the Royal Statistical Society Series B: Statistical Methodology},
  title     = {Maximum Likelihood from Incomplete Data Via the EM Algorithm},
  year      = {1977},
  issn      = {1467-9868},
  month     = sep,
  number    = {1},
  pages     = {1--22},
  volume    = {39},
  doi       = {10.1111/j.2517-6161.1977.tb01600.x},
  publisher = {Oxford University Press (OUP)},
}

@Article{Parzen1962,
  author    = {Parzen, Emanuel},
  journal   = {The Annals of Mathematical Statistics},
  title     = {On Estimation of a Probability Density Function and Mode},
  year      = {1962},
  issn      = {0003-4851},
  month     = sep,
  number    = {3},
  pages     = {1065--1076},
  volume    = {33},
  doi       = {10.1214/aoms/1177704472},
  publisher = {Institute of Mathematical Statistics},
}

@Article{Schoelkopf2001,
  author    = {Schölkopf, Bernhard and Platt, John C. and Shawe-Taylor, John and Smola, Alex J. and Williamson, Robert C.},
  journal   = {Neural Computation},
  title     = {Estimating the Support of a High-Dimensional Distribution},
  year      = {2001},
  issn      = {1530-888X},
  month     = jul,
  number    = {7},
  pages     = {1443--1471},
  volume    = {13},
  doi       = {10.1162/089976601750264965},
  publisher = {MIT Press},
}

@misc{colpi2024lisadefinitionstudyreport,
  title         = {LISA Definition Study Report},
  author        = {{LISA Definition Study Team}},
  year          = {2024},
  eprint        = {2402.07571},
  archivePrefix = {arXiv},
  primaryClass  = {astro-ph.CO},
}

\end{document}